\documentclass[11pt]{article}

\begin{document}

\title{Do "magnetars" really exist?}
\author{Malov I.F.\\
Pushchino Radio Astronomy Observatory, Lebedev Physical Institute,
RUSSIA} \date{ }
 \maketitle

 \abstract It is shown that there are neither necessary
nor sufficient properties  to provide unambiguous evidence for
including any object in the AXP/SGR class.

\section{Expected properties of AXPs and SGRs}
   To answer the question in the title we must discuss some specific properties of anomalous X-ray
pulsars (AXPs) and soft gamma-ray repeaters (SGRs). These sources
are believed  to be strongly magnetized neutron stars (magnetars)
and can be described by some additional characteristics. These
are:

1.  the supercritical dipole magnetic field $B > B_{cr} = \frac{2
\pi m^2 c^3}{e h} = 4.41 \times 10^{13}$~G,

2. low losses of the rotational energy comparing with  their X-ray
luminosities: \\
\centerline{$L_x > \frac{dE}{dt} =\frac{4 \pi I dP/dt}{p^3}$,}

3.  the bursting behaviour,

4.  the black body plus power-law X-ray spectrum,

5.  the erratic radio pulse behaviour,

6.  they are young objects connecting with SNRs,

7. they have very long periods.

~\\
Let us consider these properties one by one.
 1) Some years ago SGR0418+5729 was
discovered, with $P = 9.1$~sec (\cite{ReaEspositoTurolla10}). The
upper limit of  $dP/dt$   gives $B  = 6.4 \times 10^{19} (P
dP/dt)^{1/2} < 7.5 \times 10^{12}$~G. This object showed two
bursts at $8-200$~keV  during 20 minutes with energies $4 \times
10^{37}$ and   $2 \times 10^{37}$~ergs (the border between AXPs
and  SGRs).

Recently SGR 1822-1606 has been detected
(\cite{ReaIzraelEsposito12}). Its surface magnetic field is equal
to $2.8 \times 10^{13}$~G  and less than $B_{cr}$ as for
SGR0418+5729. So, a high surface dipolar magnetic field is not
necessarily required for magnetar-like activity. There are, on the
other hand, 19 radio pulsars with $B_s > B_{cr}$
(\cite{ManchesterHobbsTeoh05}). Hence superstrong magnetic fields
are not sufficient for the appearance of an AXP/SGR.

2) The young radio pulsar  PSR~J1846-0258 in SNR   Kes  75 ($\tau
= 884$~years) with $P = 326$~msec  shows X-ray bursts and strong
variations of times of arrivals,  i.e. it is similar to AXP/SGR .
However its losses  of rotation energy  $dE / dt = 8.1 \times
10^{36}$~erg/sec are  quite enough to provide X-ray luminosity
$L_x  = 4.1 \times 10^{34}$~erg/sec.

3) The bursting behaviour is the common characteristic of all
anomalous pulsars. However normal radio pulsars demonstrate
variations at all frequencies and at all time intervals (from
nanoseconds up to several years) as well.   Moreover giant radio
bursts of one of subpulses  are detected in a number of them (see,
for example, \cite{MalofeevMalovShchegoleva98}) and even giant
pulses are observed in some pulsars
(\cite{SoglasnovPopovBartel04}; \cite{PopovKuz'minUl'yanov06}).

Strong variability of intensity and spectral changes  of
components of individual pulses in AXP/XTE J1810-197 do not differ
in principle from the behaviour of normal radio pulsars. Their
individual pulses have not only very different intensity but their
spectral indices often changes sign at low frequencies
(\cite{Kuz'minMalofeevShitov78}). Thus anomalous pulsars differ
from radio ones by values of parameters and by the character of
their variability only.

4) Such a sum is believed typical for spectra of AXP/SGRs. However
several dozens of normal radio pulsars emit thermal and
non-thermal radiation also. Sometimes (as in the Crab pulsar PSR
B0531+21) total spectra have very complicated form. 5) There are
radio pulsars (for  example, Geminga - \cite{MalofeevMalov97})
showing changes in intensities and forms of pulses and even in
phases of pulse appearances.

6) About 20 normal  radio pulsars are observed in SNRs but  they
do not belong to the class of AXP/SGR.

7) SGR  J1627-41   has the short interval    between subsequent
observed pulses    $P =  2.6$~sec. On the other hand some
 normal radio pulsars have periods of order of several seconds
 (\cite{ManchesterHobbsTeoh05}).

 \section{Two additional arguments against the magnetar model}

1. In the popular model of magneto-rotational  explosion of
supernova (\cite{ArdeljanBisnovatyi-KoganMoiseenko05}) it is shown
that magnetic fields of order of $10^{16}$~G may only exist in a
new born neutron star for  1 sec.

2.  The detailed calculations show that  magnetic  plasmas ejected
from a neutron star emit neutrino radiation mainly.
Electromagnetic  radiation will be essential if magnetic fields in
the magnetosphere $B > 10^{16}$~G (\cite{GvozdevOgnevOsokina11}).

\section{Conclusions and discussion}
1. There are no necessary and sufficient properties to provide
unambiguous evidence for including an object in the AXP/SGR class.

2. It is not necessary to use the magnetar model for the
description of observed characteristics of AXPs and SGRs. There is
the alternative model: the drift model with the suggestion on
drift waves in the vicinity of the light cylinder
(\cite{MalovMachabeli06}). Neutron stars with rather short
rotation periods ($P < 1$~sec)  and surface magnetic fields of
order of $10^{12}$~G are believed to be the central bodies of
AXPs/SGRs in this model.

The specific characteristic of such objects is a small angle
$\beta$ between the rotation axis of the neutron star and its
magnetic moment. Indeed in those cases when radio emission of AXPs
has been detected and their polarization parameters have been
measured estimations give rather small values of angles  $\beta$.

The radio emission of two AXPs: J1810-197
(\cite{JanssenStappersKramer07} and 1E 1547.0-5408
(\cite{CamiloReynoldsJohnston08}) has shown that the variations of
the polarization position angles in these objects are small. The
maximum derivative of the position angle $\phi$ with longitude
$\Phi$ is given by

\begin{equation}
C =\left( \frac{d \phi}{d \Phi} \right)_{max} = \frac{\sin
\beta}{\sin (\zeta - \beta)} \le 1.
\end{equation}

Here, $\zeta$ is the angle between the rotational axis of the
neutron star and the line of sight toward the observer . Thus,
$\zeta - \beta$ is the minimum angular distance at which the line
of sight intersects the radiation cone. Setting the angular radius
of this cone to be $10^{\circ}$, we conclude that the angle
$\beta$ should be less than $10^{\circ}$ in J1810-197.  The
detection of an interpulse in the AXP XTE J1810-197 that is offset
from the main pulse by a distance other than $180^{\circ}$ (it is
approximately $240^{\circ}$ -- cf.\
\cite{SerylakStappersWeltevrede08}), may also directly reflect the
smallness of $\beta$ for this object.

For PSR J1642-4950 we obtain $\beta = 15.6^{\circ}$
(\cite{Malov12}). Hence, this object is also  a nearly aligned
rotator, and it is justified to apply our drift model to it.

If $\beta = 15.6^{\circ}$, the boundary of the magnetosphere is at
a distance of the order of $4 r_{LC}$, where $r_{LC}$ is the
radius of the light cylinder. This makes possible the formation of
appreciable pitch angles and the generation of synchrotron
emission, since the ratio of the magnetic energy to plasma energy
becomes less than unity. The estimates for such a case give for
AXPs/SGRs values of rotation periods $P = 16 - 250$~msec and
magnetic fields at the neutron star surface $B_s = 3.4 \times
10^{11} - 4.6 \times 10^{12}$~G (\cite{Malov10}).

In the drift model the cyclotron instability can develop near the
light cylinder,  resulting in the generation of radio emission. It
is expected that all this emission will be generated in a very
narrow layer and that it will be much more intense at low
frequencies (of the order of 100 MHz) than at higher frequencies
(\cite{Malov12}).

The main problem of all  models  is the difficulty in explaining
the energetics of power gamma-ray bursts in SGRs. Apparently,  it
is necessary to invoke sources of energy within the neutron star.
These may cause episodic ejections of plasma in the magnetosphere
and releasing of its energy, for example, as a result of nuclear
reactions (\cite{Malov12}).

{\it{\bf Acknowledgements} This work was financially supported by
the Russian Foundation for Basic Research (project code
12-02-00661) and the Basic Research Program of the Presidium of
the Russian Academy of Sciences "The Origin, Structure, and
Evolution of Objects in the Universe".}


\begin{thebibliography}{99}

\bibitem{ArdeljanBisnovatyi-KoganMoiseenko05} Ardeljan, N.,
Bisnovatyi-Kogan, G., \& Moiseenko, S.,2005, Mon. Not. R. Astron.
Soc., 359, 333.
\bibitem{CamiloReynoldsJohnston08}
Camilo, F., Reynolds, P., Johnston, et al. 2008, arXiv,
0802.0494v1

\bibitem{GvozdevOgnevOsokina11}Gvozdev, A., Ognev, I., \&
Osokina, E.2011, Astron Letters, 37, 332

\bibitem{JanssenStappersKramer07}Janssen, G.H., Stappers, B.,
  Kramer, et al. 2007, Mon. Not. R. Astron. Soc., 377, 107

\bibitem{Kuz'minMalofeevShitov78}Kuz'min, A., Malofeev, V., Shitov,
Yu. 1978,  Mon. Not. R. Astron. Soc., 185, 41

\bibitem{MalofeevMalov97}Malofeev, V., Malov, O. 1997, Nature,
389, 697

\bibitem{MalofeevMalovShchegoleva98}Malofeev, V., Malov, O.,
Shchegoleva, N. 1998, Astron. Rep., 42, 241

\bibitem{Malov10}Malov, I.
2012, Astron. Rep., 54, 925

\bibitem{Malov12}Malov, I.
2012, Astron. Rep., 56, 29

\bibitem{MalovMachabeli06}
Malov, I., Machabeli, G. 2006, Astron. Astrophys. Trans., 25, 7

\bibitem{ManchesterHobbsTeoh05}Manchester, R.N., Hobbs, G.B., Teoh, A., et
al. 2005, Astron.J., 129, 1993

\bibitem{PopovKuz'minUl'yanov06}Popov, M., Kuz'min,A.,
Ul'yanov,O., et al. 2006, Astron. Rep., 50, 562


\bibitem{ReaEspositoTurolla10} Rea, N., Esposito, P., Turolla, R., et al.
2010, Science, 330, 994

\bibitem{ReaIzraelEsposito12} Rea, N., Izrael, G., Esposito, P., et al.
2012, arXiv, 1203.6449v1

\bibitem{SerylakStappersWeltevrede08} Serylak,M., Stappers, B.,Weltevrede,
P., et al. 2008, arXiv 0811.3829v1

\bibitem{SoglasnovPopovBartel04}Soglasnov, V., Popov, M., Bartel, N., et
al. 2004, Astrophys.J., 616, 439


\end{thebibliography}
\end{document}